\documentclass{article}
\usepackage[disable]{todonotes}

\usepackage{ijcai21} 
\usepackage[authoryear,square]{natbib}

\usepackage{times}
\usepackage{adjustbox}
\usepackage{soul}
\usepackage{url}
\usepackage[hidelinks]{hyperref}
\usepackage[utf8]{inputenc}
\usepackage[small]{caption}
\usepackage{graphicx}
\usepackage{amsmath}
\usepackage{amsthm}
\usepackage{booktabs}
\usepackage{algorithm}
\usepackage{algorithmic}
\title{AI video editing tools \\ {\small What editors want and how far is AI from delivering?}}

\author{
    Than Htut Soe
    \affiliations
    University of Bergen
    \emails
    than.soe@uib.no
}

\begin{document}
 
 \maketitle

\begin{abstract}
Video editing can be a very tedious task, so unsurprisingly Artificial Intelligence has been increasingly used to streamline the workflow or automate away tedious tasks. However, it is very difficult to get an overview of what intelligent video editing tools are in the research literature and needs for automation from the video editors. So, we identified the field of intelligent video editing tools in research, and we survey the opinions of professional video editors. We have also summarized current state of the art in artificial intelligence research with the intention of identifying what are the possibilities and current technical limits towards truly intelligent video editing tools. The findings contribute towards understanding of the field of intelligent video editing tools, highlights unaddressed automation needs by the survey and provides general suggestions for further research in intelligent video editing tools.
\end{abstract}




\section{Introduction}\label{sec:introduction}


Video is the most popular form of content on the Internet. According to the Cisco visual networking index ~\citep{cisco_cisco_2018}, 75\% of the Internet traffic in 2017 has been video content. Mobile phones, video sharing and social media platforms make it easier and quicker than ever to capture and publish videos. Editing those videos, however, is still very time-consuming. Video remains a difficult medium to edit as it requires operation at individual frames on top of being a dual-track medium with both audio and image. There are various attempts to make video editing easier. One approach is to automate the editing process using artificial intelligence (AI). We are here interested in the state of the art in video editing automation, specifically focusing on the discrepancy between what is desired and what is attainable by current AI technology.


Okun et al. \citep{okun_ves_2015} define video editing as the act of cutting and joining pieces of one or more sources together to make one edited movie. \emph{Video editing tools} can roughly be defined as  (computer) programs  that people can use to do the task of video editing, i.e., combining video segments. Video editing is one of the areas where AI has been used to automate or augment the tasks of human video editors. 

\emph{Intelligent video editing tools} have been attempted since the beginning of digital video editing with the goal of making video editing easier. One of the key themes in intelligent video editing tools is the problem of enabling the manipulation of the video from a high level of abstractions, for example shots and dialogue rather than frames. One early example of such a tool is    \emph{Silver} \citep{casares_simplifying_nodate} from 2002, which provides smart selections of video clips, as well as abstract views of video editing, by using metadata on from the videos. A more recent example of an intelligent video editing tool is \emph{Roughcut} \citep{leake_computational_2017}. \emph{Roughcut} allows the computational editing for dialog driven scenes using user input of dialog for the scene, raw recordings, and editing idioms. There is an open source tool \emph{autoEdit} \citep{passarelli_autoedit_2019} and a research \citep{berthouzoz_tools_nodate}, which enable text-based editing of video interviews by linking text transcripts to the videos. \todo {Not clear what "to be edited with text" means?}

Entirely AI controlled video production has received a lot of research interest lately \citep{xue_automated_nodate,hua_optimization-based_2004}. At present, AI controlled video production is aimed at creating automated {\em video summaries} or {\em mashups}. 
These completely automated video editing methods, as used to create video summaries and mashups,  are not be considered as intelligent video editing tools because they are simple algorithms that execute a very narrow and specific request that requires no intelligence or user interaction involved.  


Advances in AI in image processing, vision and natural language processing have made numerous automation and augmentations in video editing possible. But, has the dream of shrieking the drudgery of video editing been accomplished? The answer, of course, hinges on whose dream we are talking about. How can progress in automated video editing be evaluated? This question of evaluating the progress in automatic video editing can be approached from two angles. The first is to conduct an overview of literature, while the second is to survey the expectations from human video editors and line them against the state of the art in AI. In this paper we do both.  

The main challenge that intelligent video editing tools tried to solve was the streamlining of the video editing process for the users. This is usually done by removing the tedious tasks of looking through video clips frame by frame. Solutions and tools proposed differ in a verity of ways that stem from the approach to the problem, the intended purpose, the underlying technology, the level of abstraction(s), the interactions offered, and the modalities for the said interactions. We conducted a review into the state of the art in  video AI applications from two perspectives: i) general AI technology for video; ii) video editing specific AI technology. General AI technology for video includes a wide variety of video tasks such as object tracking, object detection, speech recognition, video reasoning, action detection, sentiment detection in videos. Specific video editing AI technology is much narrower such as processing video scripts, shots and scenes and mining video editing rules. 


We conducted a survey of 13 video editors whose video editing experiences ranges from 1 year to 22 years. The survey covers their background in video editing, thoughts on AI video editor and automation needs in video editing. The responses from the survey are then used to perform thematic analysis to form an overview of expectations, requirements, and issues regarding automation in video editing tools. We compare the opinions and expectations uncovered from the survey with current knowledge about machine learning for content creation/manipulation, automated video editing, and other AI tools. We discuss how state of the art in AI can   create an ideal AI video editing tool for video creators.

This paper is structured as follows: Section~\ref{sec:background} provides an overview of intelligent video editing tools and AI techniques in video. Intelligent video editing tools in literature are compared and summarized in Section~\ref{sec:intelligenteditors}. The survey of  (human) video editors, including the procedure and summarized results are in Section~\ref{sec:survey}. In Section~\ref{sec:challenges}, previous works on solutions for intelligent video editing tools and that of the expectations of users from the survey is contrasted and some AI techniques are proposed as potential solution to meet the expectations from the users. Finally, the summary of the paper, conclusions and future work is presented in the Section~\ref{sec:conclusion}.

\section{Background}\label{sec:background}

Creating better tools to make video editing easier has always been a research agenda since the introduction of digital video. We first given an overview of different approaches to creating intelligent and automatic video edition tools. We then turn to the AI methods that have been applied directly or indirectly in video editing.

One of the first projects to attempt to simplify video editing is \emph{Silver}\citep{myers_multi-view_2001, long_video_nodate}. In the first version of the project, Silver has different types of views, namely, transcripts, timeline, preview, and storyboard views. This editing tool also explores intelligent editing with smart selection, cut, delete, copy, paste, and reattach using shots and scene detection. Intelligent editing is built using the metadata layer of the video in the form of text transcript, short boundaries detection, and Optical Character Recognition (OCR) to generate metadata. In the second iteration of the Silver tool \citep{long_video_nodate} the tool implemented lenses (clip, shots, frames) and semantic zooming to make easier tasks such as visualizing the right frames to cut when joining two video segments. Most intelligent video tools are created for making just one particular type of video. For example, QuickCut \citep{truong_quickcut_2016} is created for composing narrated videos and  Video digests \citep{pavel_video_2014} is created for summarizing lecture video.



Next, we consider methods for completely automated video editing. \todo{What is completely automated video editing. Give a definition} Automated video editing is  computationally processing and composing video segments without any input from a human editor. Automated video editing can be performed on recorded video clips or a much larger scale on video archives. Early works on automated video editing focus on rule-based video sequence generation strategies \citep{butler_film_1997} or semantics-based method for selection and automated editing of user requests in domain of video documentaries \citep{bocconi_semantic-aware_2004}. Mashups, combination multiple video clips about a single event, is another type of automated video editing. The work named Virtual Director \citep{shrestha_automatic_2010} created a mash up generation method for concert recordings which maximizes what makes a good concert video based on rules sourced from interviewing video editors and film grammar literature. Automated video editing can also be used to broadcast live events. Work by \citep{radut_how_2020} discussed not only the prototype AI video editor for live events but also an evaluation and discussions of evaluation methods for measuring the quality of AI edited live events. 
AI can be used to automatically create edited videos as well. {\em Made by Machine When AI met the Archive from BBC} created 150 short complications from the BBC archive    \citep{bbc_rd_ai_2018}. \citep{taskir_automated_2006} presented a summation method for skimming of video programs using speech transcript from speech recognition systems. \citep{truong_quickcut_2016} provides a summary of video abstractions or summation methods such as generating a sequence of keyframes or moving images for the purpose of providing information about a video in a shortest period of time possible. The work on automated video editing of corporate meetings by \citep{wu_joint_2020} uses learned editing decisions from human edited video and uses two attention models on both audio and video. \textsc{EDL}-Edit Decision List is a text-based language that encodes composition decisions with ordered list of clips and time code data. It is used as the output of many automated video editing systems \citep{taskir_automated_2006, wu_joint_2020, passarelli_autoedit_2019} which the video editors can then use the EDL encoded automated editing decisions to further continue their editing in software such as Adobe Premier Pro or Davinci Resolve.





How AI techniques can be used to extract information from videos is a very diverse area of research. We are particularly interested in facial recognition, object detection, object tracking, scene detection, sentiment analysis, video reasoning and video captioning. \todo{You need one sentence to define what each of these means. I did the ones I can do with confidence} Facial recognition refers to the problem of identifying whether a human face is present in an image, and possibly whose, while object detection is the problem of identifying a specific object in an image. Object tracking is the problem of identifying and locating a specific object and tracking its movement across the frames in a video. Scene detection, or video segmentation, is identifying segments which are semantically or visually related in a video. Sentiment analysis is the problem of matching the sentiment that would be conveyed by a given content: is it happy, sad, ironic etc. Video captioning \citep{wu_deep_2016} is an AI technique that generates natural descriptions that capture the dynamics of the video.

\todo{find scene/cut detection}




There are video techniques from AI that are more specific to video editing. \citep{matsuo_mining_nodate} presented data mining technique to discover editing patterns (made up of loose, medium tight shots and rules) from videos with the goal of creating reproducible editing patterns. Earlier works by \citep{butler_film_1997} presented rules and query-based approach to automate video editing. Automated video editing by modelling the editing process and using semantics is presented in \citep{zhang_application_1997}.





\section{Intelligent video editing tools}\label{sec:intelligenteditors}
\todo{ add this somewhere \citep{ueda_automatic_1993} structure visualisation} 

In this chapter, we do a literature review and summarize how previous intelligent video editing tools formulate and solve the problem of making video editing an easier task. The literature search for intelligent video editing tools is performed with the keywords \emph{(intelligent OR smart OR automated OR AI) AND  (video editor OR video editing)} in computer science literature databases which are  DBLP, ACM digital library and Google Scholar. Titles and abstracts were then read and filtered based on the inclusion criteria stipulating that included papers must describe an intelligent approach to making a video editing tool for users. Included papers had to also contain a description and/or implementation of the user interface. The references of the included papers were also scanned to discover more related literature. The resulting papers were then summarized and grouped in terms of three topics, namely, \emph{video editing tasks}, \emph{interaction with automated editor} (human computer interaction) and \emph{AI technology}. The list of included papers in this section is in Table~\ref{table:paperslist}
\begin{table}
\begin{adjustbox}{width=\columnwidth,center}
\begin{tabular}{p{3cm}|p{3cm}|p{8cm}} \\
Work & Video type & Goal \\
\hline
\citep{casares_simplifying_nodate}	& edited videos &	video editing more accessible to novices, make video editing as easy as text \\
\citep{long_video_nodate} 	& 	edited videos &	Scale and zooming with multiple lenses - new interactions \\
\citep{leake_computational_2017}	& 	dialog driven scenes &	efficiently explore the space of possible edits \\
\citep{kimura_video_2005}		& any video	 & Creating semi-edited videos from gaze data \\
\citep{shipman_authoring_nodate}		& any video	 & detail-on-demand" interactive video  summaries because searching for information in the video is difficult \\
\citep{chi_democut_2013}	& 	 how to videos  &	streamline amateur editing of how-to demonstration videos by "semi-auto" editing \\
\citep{berthouzoz_tools_nodate} 	& 	interview videos   &	to make the task of making interview video easier \\
\citep{truong_quickcut_2016}	& 	narrated videos	 & interactive video editing for efficiently log raw footage and editing narrated video \\
\citep{pavel_video_2014}	& 	informative videos	 & Make the long informative videos easier to browse \\
\citep{cattelan_watch-and-comment_2008}	& any video &	easy way to create videos for end users watch and comment paradigm  \\
\hline
\end{tabular}
\end{adjustbox}
\caption{List of papers included in the study} \label{table:paperslist}
\end{table}



\subsection{Video editing tasks - of intelligent video editors} 
This subsection presents  the field of intelligent video editing tools in terms of different tasks addressed in the video editing workflow and summarizes the approach for each task. 
\todo{may be a table to summarize this section}

\emph{Segmentation of videos} is the most common task intelligent video editors try to solve. All of the previous work reviewed in this section used a certain form of video segmentation method, but different approaches are used to perform the segmentation. From now on a video clip is defined as a continuous segment of video from a single source file. The first segmentation method identified is using shot detection. A shot is an unbroken continuous image sequence \cite{okun_ves_2015}. Segmentation with shot detection is performed with image analysis methods in \citep{casares_simplifying_nodate, long_video_nodate} and shot detection is done via features which are camera motion, brightness and duration in \citep{shipman_authoring_nodate}.  \citep{wu_movieup_2015} uses shot and sub-shots for segmenting user generated videos (subshot is defined as basic unit of video which contains consistent camera motion and self-contained semantics). \citep{casares_simplifying_nodate} also considered segmentation of video and audio at different locations in case of L-cuts.  

The second segmentation method uses synchronization with text transcript for creating cuts which depend on content or the meanings from the audio track of the video. \citep{leake_computational_2017} used pre-written script lines of the dialog of the scenes to segment the video while \citep{pavel_video_2014} passed the text transcript through Bayesian topic segmentation\citep{eisenstein_bayesian_2008} BSec to identifying sections and subsections of informational videos. Other approaches to segmentation include approaches such as using gazes \citep{kimura_video_2005}, starting from user marked points, frame similarity measures with those points \citep{chi_democut_2013} and cut suitability score in interview videos with a talking head \citep{berthouzoz_tools_nodate}. User generated video summaries by \citep{cattelan_watch-and-comment_2008} created segmentation using user watch actions and comments. Segmentation of video is also essential for discussing the next task which is composition of video segments. 

\emph{Composition of video segments} is the second most common task addressed in our list of literature on AI video editing tools. The most common approach to streamline composition of video segments is using dialogues in the scene \citep{leake_computational_2017} or text transcripts \citep{berthouzoz_tools_nodate, wu_movieup_2015} as a starting point of the composition. Dialogues of the video must be written and provided as an input, but text transcript can be generated using speech recognition technology. For example, \citep{truong_quickcut_2016} uses text transcript converted from narrated audio or voice-overs instead of manually created text transcripts. In Roughcut\citep{leake_computational_2017}, the video segments created for each corresponding dialogue and speaker using automation. The order of the dialogue follows the provided script. However, the creative composition can be changed by the user selecting combination of video editing idioms. Similarly, editor created story outline had been used to compose  segments\citep{truong_quickcut_2016}.  Composing segments is done by means of cutting out unwanted parts, such as certain phrases of an interview or repeated words, by selecting the corresponding text from the video transcript \citep{berthouzoz_tools_nodate}.

\citep{chi_democut_2013} uses user provided markers in the video clips as a way to organize and compose segments and editors can change the composition of the overall segment by arranging the markers which correspond  to steps in a demonstration video. Composition for the purpose of creating user generated video summaries is done using modelled viewer intent from gazes \citep{kimura_video_2005} and with watching activities and user comments in \citep{cattelan_watch-and-comment_2008}.

\emph{Visualization of the timeline and video clips} comes in the form of viewing the timeline and video clips at different levels of abstractions and different representing the video timeline in alternative fashion such as text.  Abstractions of visualising a video can be in the form frame, shot and clip. A frame is a still image of a video while a shot is an unbroken continuous image sequence \citep{okun_ves_2015}. A review of different forms of video abstractions is available in \citep{truong_video_nodate}. Visualization of the clips using a representative frame is discussed as a method to allow quick judgement of the content of the video on the timeline \citep{long_video_nodate}. \citep{casares_simplifying_nodate} proposed visualization of the timeline in different levels of abstractions which are storyboard, editable transcript and timeline views. The second approach to visualization of the timeline is via representation of the timeline in terms of text transcripts \citep{casares_simplifying_nodate, truong_quickcut_2016,berthouzoz_tools_nodate, pavel_video_2014}. As noted in the previous paragraphs, textual representation of the timeline can sometimes be manipulated on word level in order to make changes in the actual composition of the video frames. 

\emph{Smart manipulation of clips} is only discussed in two of the intelligent video editors. The first work \citep{casares_simplifying_nodate} used smart selection, snap, cut, paste, and reattach. All of these actions are done using shot boundaries  with image analysis. The second work just provides smart selection or smart cutting of video segments using the transcript of the video \citep{berthouzoz_tools_nodate}. The lack of many examples for this task and the tasks mentioned below might have to do with the fact that all intelligent video editing tools are proof of concepts thus lacking these very important, but non essential features. 

\emph{Creating transitions.} Easier method  for creating transitions is addressed in two separate works. \citep{truong_quickcut_2016} created a method for automating aesthetically pleasing transition by formulating transition  tasks as  dynamic programming in which bad transition  points such as jumpcuts are penalized. \citep{berthouzoz_tools_nodate} uses a different approach to create hidden transitions by using hierarchical clustering of frames and finding out the shortest path between frames as transaction points.  

\emph{Logging of videos.} \citep{truong_quickcut_2016} presented a novel approach to logging of video clips with audio annotations by logging of video during the filming process. In their work, logging can be done via audio, in addition to logging with tags during the review of the footage.


\subsection{Interaction with automation}
In this section, the mode of interaction as well as the level of video abstractions \citep{truong_video_nodate}   will be summarized. The primary mode of interaction used in most of the intelligent video editing tools we explored is via Graphical User Interface (GUI) with a keyboard and mouse. An  exception is the one case by \citep{kimura_video_2005} which explored a gaze-based interaction. However, the level of abstraction and granularity of control that the users have is different in different tools. In video editing tools without any abstraction, editing must be done at the level of individual frames which is very labor intensive. However, some of the intelligent video editing tools offer  manipulation of the video at multiple abstraction  levels. Two examples of tools offering  multiple abstractions are Silver\citep{casares_simplifying_nodate} and Quickcut \citep{truong_quickcut_2016}. Silver which offers three abstractions which are: clip, shot and frames. Quickcut offers abstractions in terms of spoken words and frames. 

Some video editing tools work  at a very high level of abstraction. In DemoCut\citep{chi_democut_2013} for example, videos editing takes place at the abstraction of steps and markers for these steps. Similarly in RoughCut\citep{leake_computational_2017}, the user can manipulate the timeline using dialog lines in a dialog script and editing decisions in the form of idioms. Manipulation at   higher levels, however, comes at the cost of the ability to make finer adjustments at the frames level. However, in three of the intelligent video editing tools \citep{leake_computational_2017,truong_quickcut_2016,passarelli_autoedit_2019} video editing work can be exported as EDL (Edit Description Language) which can be used with other commercial video editing software in order to do adjustments at frames level and complete the video editing process. 




\subsection{AI technology being employed}
\emph{Video Segmentation.} Earlier work on intelligent video editing tools relies on image analysis for detecting shot boundaries and finding representative frames for each shot \citep{casares_simplifying_nodate} or using a combination of image analysis, domain knowledge and model matching in \citep{shipman_authoring_nodate}. The rules for detecting shots are handcrafted for targeted video types. In \citep{casares_simplifying_nodate}, hand crafted transcripts are aligned to videos using speech recognition. Segmenting video lectures into topically-coherent units is done by performing topic segmentation on the text transcript of the video in \citep{pavel_video_2014}. Another form of segmentation with audio annotations is explored in \citep{truong_quickcut_2016}. It works by employing motion based segmentation and refining via audio annotation into semantically relevant segments. Motion-based segmentation is done by detecting continuous motion in the video while semantic segmentation corresponds to actions or topics in the video. 

\emph{Representation of filming principles} using computational techniques can be found in \citep{wu_movieup_2015} and \citep{leake_computational_2017}. Domain specific principles for detecting   video cut points, selected video shots  and selected audio fragments, are curated through interviews and represented as optimization problems \citep{wu_movieup_2015}. In \citep{leake_computational_2017}, 12 basic film editing idioms (avoid jump cuts, intensify emotions, etc.) are represented in terms of feature parameters that go as input into a hidden Markov model for generating editing decisions. A Hidden Markov Model (HMM) is a statistical approach for modelling sequences in which the series of internal states are hidden. The features used in the HMM includes labels generated using speech to text, face detection and structural information from the clips. 






\section{What do editors want from an intelligent video editing assistant}\label{sec:survey}

In this section we report on the survey we conducted to explore the opinions of (human) video editors regarding what constitutes an ideal AI video editor.  

\subsection{Survey procedure}
For the survey we considered not only video editors from the broadcasting industry, but also independent video editors. The survey was sent out via email to a a list of video editors affiliated to a broadcasting company. To get the opinions of independent video editors, the survey was posted as a task to independent video editors on upwork.com with a 10 USD response reward. We received responses from 13 participants in total, 5 from the broadcasting industry and 8 from among the independent video editors.  The survey consisted of 12 short questions organised in three topics: background questions on video editing experience and knowledge of AI, what their ideal AI editor is,  and what do they need from automation in video editing. 
The full survey question and anonymous responses are available in <link removed for review>



\subsection{Survey results}
 

\emph{Background information.} The average years of experience in video editing among our participants in the survey is 9.75 years with the shortest being 1 year to longest being 22 years. In terms of the type of video the participants work with, each participant listed around 3 types of video each. The most common video types are commercial, documentaries, presentation, sports, social media videos and news. 

In terms of software programs, the participants have used 5 programs on average in video editing. The most frequently mentioned  editing programs are Adobe Premier Pro and DaVinci resolve. In addition, lesser-known programs such as Avid, Flimora 9, and VizStory are also mentioned, as well as video services such as Rev.com and Descript.com. For the question probing which AI technology have they  heard about in the context of video editing,  8 out of 13 respondent answers with AI technology in branded and commercial offering such as Adobe Premier Pro CC and Magisto. For the remaining 5 respondent the answers cover AI techniques which are auto correction, noise reduction, background removal, video stabilization, objection detection/image annotation, segmentation, up-scaling, deep fake, automated video digests, face recognition and speech to text. 

\emph{Ideal AI editor}. The response to the question \emph{``What would you like the perfect AI video editing tool be?''} the answers differ  significantly from one another. However, we identified five themes in the answers. They are  AI tool for video editing tasks, AI as a tool for project management tasks, automatic aesthetic quality improvement tools, human-AI concerns, and AI for content discovery.

AI as a tool for video editing tasks is the largest category that most responses fall  into. This contains keywords which are shot detection, composition of clips, filtering bad video takes based on dialogue lines, synchronization of tracks and subtitles, translation, and language understanding. The AI for project management theme includes terms which are video metadata creation, data management, and ingest. The next theme is AI for aesthetic quality improvement which covers automatic color grading and automatic audio equalization. In addition to that there are human-AI concerns with regards to AI such as balance of control and automation, user centered AI, and personalization. Lastly, the terms in the AI for content discovery theme include suggesting stock video footage and stock music video based on existing content on the timeline. 

On the question regarding the mode of interaction with the AI video editing tool, most of the responses mentioned that they would like to interact via voice followed by those who would like to interact via  Graphical User Interface with keyboard and mouse. In addition to these two major interaction modes, various other modes such as touch interface, gestures, brain computer interface are also mentioned a few times. Some responses mentioned contextual commands based on the state of the project as essential for communication, as well as an AI OFF button what allows the automation to be shut off easily. 

The last question on the perfect AI editor topic asks for the level of abstractions the (human) editors would like to work with. Most of the respondents said that they would like to manipulate the video at keyframes level in their vision of perfect AI video editing tool. The second and third most popular level of abstraction are clips and frames. Other types of abstractions which are mentioned once each are sequences, story and shot. Two respondents mentioned that they would like a flexible type of abstraction where they can adjust the level of abstractions per interaction basics.

\emph{AI and workflow.} In the second part the survey, the following topics are explored: tasks in video editing workflow the participants want to automate, and related questions with level of autonomy and interaction modes for these tasks. The responses to the questions in the workflow part consist of four thematic tasks: video editing tasks, aesthetic improvements, video pre-editing tasks, and suggestive tasks.

The most popular keywords used to describe the video editing tasks are composition of segments and subtitling which are mentioned in three responses. Second most popular video editing tasks keywords are video segmentation and filtering out bad shots (which are mentioned twice). In addition, the following video editing tasks are mentioned once: content analysis, video archival, facial recognition, placing cuts, choosing transition frames,  synchronization of audio to video, and dual track audio video selection. 

The next frequently mentioned thematic task in the would-like-to-automate responses is aesthetic quality improvements.  Most popular terms for these are color correction and audio equalization. In addition, tasks like visual improvements, background removal and stuttering removal were mentioned once each. For the pre-editing tasks, logging of videos is mentioned twice, and automated creation of time code is written once. In terms of suggestive tasks, the responses includes clip suggestion with editing styles, general assistance and music suggestions.

The last question is  how similar the editors expect the AI editor  to be in comparisons to the tools they have been using. In that question four respondent said that they want it to be very similar or familiar. Two responses said they want it to have a basic level of similarity. The keyword ``easy to use'' found in two responses is another word that might have similar meaning to ``familiar''. One respondent mentioned a plugin approach for integrating AI video editing tools inside existing tools. Only two respondents said they expect the AI video editors in the future to be very different or not similar at all.

\section{Challenges for AI video editing tools}\label{sec:challenges}

\emph{Video editing tasks}  There is a significant overlap between video editing tasks identified in the literature and mentioned in our survey results. We focus here on the unexplored parts of the video editing workflow. The first is the  synchronization of audio and videos from different tracks. This task has already been explored in a different but related context of automated video mashups generation \citep{wu_movieup_2015,shrestha_automatic_2010}. Filtering out bad takes or bad segments in video has not been explored but this has to be done with  informed research on understanding of what do the video editors meant when they said bad video takes. Lastly, we have language issues such as automated translation, subtitling and language understanding in video editing. The subtitling and automated translation system can be automated using machine automation. However, the language understanding in the context of video editing requires both advances in natural language processing and understanding the usages of video editing terms and language in video editing context. 


\emph{Video logging and metadata creation} Video logging is watching a video footage and labelling its contents by time codes. It has been identified in both the intelligent video editing literature and our survey as one of the tasks that the users would like to automate. Current techniques for video logging in the literature are very limited to specific applications; namely, demonstration videos and dialogue-based videos. Speech recognition to convert the video into text and apply text processing techniques has a lot of potential use cases in video logging and metadata creation. Another interesting AI area to watch out for is video reasoning and understanding  combinations of patterns from both the visual and the language input \citep{wu_deep_2016}. 


\emph{Voice-based video editing interactions} is the most common mode of interaction people said they would like to use to communicate with an AI video editing tool. The potential of Voice User Interfaces in video editing tools has not been explored. \citep{chang_how_2019} presented the exploration of the design space of voice-based interactions for navigation of how-to videos. Since voice interaction in video editing is entirely unexplored territory, the starting point must be a design exploration. Another possibility is exploration of single voice commands without context, instead of interactions. Tasks that the user would like to fully automate, such as  aesthetic quality improvements, file management or pre-editing tasks, should be ideal for designing voice commands. 


\emph{Personalization} in  this context is the ability of an intelligent video editing tool to adapt to a user by learning from videos processed, video produced and usage patterns in the software. This topic is absent in the literature, but it is found in our survey in the form of understanding the context of the video editing and personalization. Rule-based learning of video editing rules is discussed in  \citep{matsuo_mining_nodate}. 





It is important to note that the intelligent video editing tools discussed in the Section~\ref{sec:intelligenteditors} are created to solve video editing of a particular type of video: in fact, eight of them target only a single type of video. Our survey of video editors however, listed three types of videos handled by each on average. Since most of the intelligent video editing tools are created for a single type of video, there is a concern about the general applicability of techniques mentioned in the results.

Video editing tasks described in the intelligent video editing tool literature are focused on interactions during the core video editing task. The survey participants however have automation needs for additional tasks such as file/media organization, aesthetic quality improvements, pre-editing tasks and content suggestion. Content suggestion for video editing  is suggesting good video clips or music segments to add to an existing video or story. 

In the survey results, voice interaction is the most common mode of interaction wanted by the participants. However, only one work on intelligent video editor included voice \citep{chi_democut_2013} where voice annotations are used to tag videos. This popularity of voice interaction might be due to the popularity of AI based voice assistance in mobile phones and smart home speakers as well as portrayal of AI as voice in the science fiction. 



The AI techniques used in the intelligent video editing tools are heuristics-based systems. Other approaches such as neural network and machine learning based approaches are less explored. The study of \citep{dove_ux_2017} concludes that machine learning is a difficult design material to work with for creating user experience, as it is difficult to create prototypes based on machine learning and it requires collaborators from machine learning to execute it.  





\section{Conclusion}\label{sec:conclusion}

In this paper we have defined intelligent video editing tools and presented a review of the existing literature regarding (intelligent) video editing, user interaction and AI technology. We have also surveyed video editors regarding their needs for automation in their video editing workflow. Doing research in this field of study requires knowledge on video editing, human computer interaction and AI or machine learning. Intelligent video editing tools requiring these three very different expertise a is one of the reasons the literature on it is very limited compared to each of the fields on its own. 

There is a good amount of crossover between the literature and survey results in this study in the area of video editing tasks. However, there are areas such as logging of videos, organization of video editing projects, aesthetic quality adjustment and content suggestion that need to be explored further to fulfill the needs of from the survey. It is our conclusion that with a greater involvement of the machine learning community, the ideal AI editor can be reached. In future work we intend to contribute towards advancing this goal.

\bibliographystyle{abbrnatths}
\bibliography{1_ideal_ai_editor}

\end{document}